\documentclass[letterpaper,floatfix,aps,pra,amsmath,amssymb,twocolumn,superscriptaddress,
showpacs,footinbib]{revtex4-1}

\usepackage{epsfig}
\usepackage{verbatim}
\usepackage{color}
\usepackage{epsf}
\usepackage{graphicx}
\usepackage{array}
\usepackage[pdfauthor={Riwar, Hosseinkhani, Burkhart, Gao, Schoelkopf, Glazman, Catelani},
            pdftitle={Normal-metal quasiparticle traps for superconducting qubits}]{hyperref}

\hypersetup{pdfstartview={XYZ null null 1.09}}

\newcommand{\be}{\begin{equation}}
\newcommand{\ee}{\end{equation}}
\newcommand{\bea}{\begin{eqnarray}}
\newcommand{\eea}{\end{eqnarray}}

\newcommand{\eref}[1]{Eq.~(\ref{#1})}
\newcommand{\esref}[1]{Eqs.~(\ref{#1})}
\newcommand{\rref}[1]{(\ref{#1})}

\newcommand{\ocite}[1]{Ref.~\onlinecite{#1}}

\newcommand{\qp}{\mathrm{qp}}

\renewcommand\[{\begin{equation}}

\renewcommand\]{\end{equation}}

\begin{document}

\title{Normal-metal quasiparticle traps for superconducting qubits}

\author{R.-P. Riwar}

\affiliation{Peter Gr\"unberg Institut (PGI-2) and JARA Institute for Quantum Information, Forschungszentrum J\"ulich, 52425 J\"ulich, Germany}
\affiliation{Departments of Physics and Applied Physics, Yale University, New Haven, CT 06520, USA}

\author{A. Hosseinkhani}

\affiliation{Peter Gr\"unberg Institut (PGI-2) and JARA Institute for Quantum Information, Forschungszentrum J\"ulich, 52425 J\"ulich, Germany}
\affiliation{JARA-Institute for Quantum Information, RWTH Aachen University, D-52056 Aachen, Germany}

\author{L.~D. Burkhart}

\affiliation{Departments of Physics and Applied Physics, Yale University, New Haven, CT 06520, USA}

\author{Y.~Y. Gao}

\affiliation{Departments of Physics and Applied Physics, Yale University, New Haven, CT 06520, USA}

\author{R.~J. Schoelkopf}

\affiliation{Departments of Physics and Applied Physics, Yale University, New Haven, CT 06520, USA}

\author{L.~I. Glazman}

\affiliation{Departments of Physics and Applied Physics, Yale University, New Haven, CT 06520, USA}

\author{G. Catelani}

\affiliation{Peter Gr\"unberg Institut (PGI-2) and JARA Institute for Quantum Information, Forschungszentrum J\"ulich, 52425 J\"ulich, Germany}

\begin{abstract}
The presence of quasiparticles in superconducting qubits emerges as an intrinsic constraint on their coherence. While it is difficult to prevent the generation of quasiparticles, keeping them away from active elements of the qubit provides a viable way of improving the device performance. Here we develop theoretically and validate experimentally a model for the effect of a single small trap on the dynamics of the excess quasiparticles injected  in a transmon-type qubit. The model allows one to evaluate the time it takes to evacuate the injected quasiparticles from the transmon as a function of trap parameters. With the increase of the trap size, this time decreases monotonically, saturating at the level determined by the quasiparticles diffusion constant and the qubit geometry. We determine the characteristic trap size needed for the relaxation time to approach that saturation value.
\end{abstract}

\date{\today}


\maketitle

\section{Introduction}
\label{sec:intro}

Ideal superconducting devices rely on dissipationless tunneling of Cooper pairs across a Josephson junction. For example, in a Cooper pair pump \cite{cpp}, the controlled transport of Cooper pairs across two or more junctions can in principle make it possible to relate frequency and current and hence enable metrological applications of such a device \cite{cppmet}. For quantum information purposes, the non-linear relation between the supercurrent and the phase difference across a junction makes the junction an ideal non-linear element to build a qubit \cite{sqr}. However, in addition to the pairs tunneling, single-particle excitations known as quasiparticles can also tunnel. In the pumps this leads to ``counting errors'', limiting the accuracy of the current-frequency relation \cite{cpp,cppmet}. In qubits,
quasiparticles interact with the phase degree of freedom, providing an unwanted channel for the qubit energy relaxation~\cite{prl,3dtr}. While in many cases it is impossible to prevent the creation of quasiparticles, one may keep them away from the Josephson junctions by trapping. Evacuation of the quasiparticles from the vicinity of the junction  provides a way to extend the energy relaxation time ($T_1$) in the steady state, and to restore the steady state after a perturbation, whether caused by qubit operation or some uncontrolled environmental effect.

Quasiparticle trapping has been explored for a long time, and various proposal exists on how to implement such a trapping. For example, gap engineering takes advantage of the fact that quasiparticles accumulate in regions of lower gap to steer them into or away from certain parts of the device. Gap engineering was used successfully to limit quasiparticle ``poisoning'' in a Cooper pair transistor \cite{cpt}, while proved ineffective in a transmon qubit \cite{sun}.
A vortex in a superconducting film can also act as a well-localized trap, since the gap is completely suppressed at the vortex position. Trapping by vortices has been demonstrated \cite{ullom,plourde,wang,pekola2}, but vortex motion may induce an unwanted dissipation. An island of a normal metal in contact with the superconductor may also serve as a quasiparticles trap~\cite{raja1,raja2}. In the limit of weak electron tunneling across the contact, the proximity effect~is negligible. The quasiparticles tunneled into the normal metal are trapped there upon losing their energy~by phonon emission or inelastic electron-electron scattering.

The majority of previous works concentrated on the control of a steady-state quasiparticle population~\cite{cppmet,raja1,raja2}. In contrast, we are interested in the effect of a normal-metal trap on the dynamics of the quasiparticle density. Traps accelerate the evacuation of the excess quasiparticles injected in a qubit in the process of its operation. Our main goal is to determine how the characteristic time of the evacuation depends on the parameters of a small normal-metal island in contact with the superconducting qubit. The characteristic time shortens with the increase of the trap size, saturating at a value dependent on the qubit geometry and the quasiparticle diffusion coefficient. The size at which a trap becomes effective depends on the contact resistance, the energy relaxation rate in the normal-metal island, and the effective temperature of the quasiparticles. We develop a simple model allowing to evaluate the time evolution of the quasiparticle density and find the characteristic evacuation time as a function of the trap parameters. The model is validated by measurements of the qubit $T_1$ relaxation time performed on a series of transmons with normal-metal traps of various sizes.

The paper is organized as follows: in Sec.~\ref{sec:diff} we develop a phenomenological quasiparticle diffusion  and trapping model which includes the effect of a normal-metal trap. In Sec.~\ref{sec:dyn} we study the dynamics of the density during injection and trapping in a simple configuration, and in Sec.~\ref{sec:exp} we provide experimental data supporting our approach.
We summarize the present work in Sec.~\ref{sec:summ}.

\section{The diffusion and trapping model}
\label{sec:diff}

\begin{figure}[t]
\begin{center}
\includegraphics[width=0.42\textwidth]{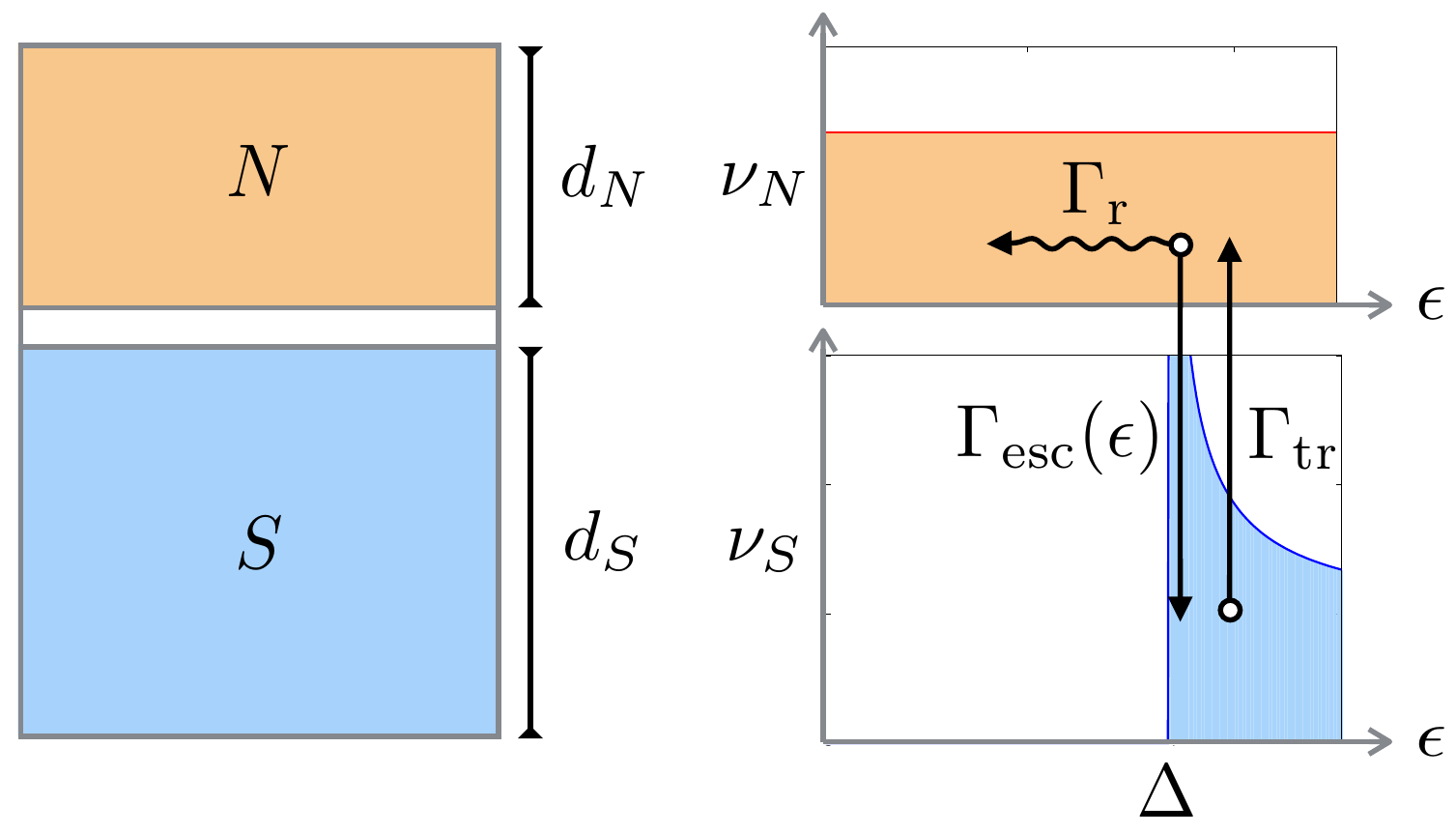}
\end{center}
\caption{Left: a small superconductor $S$ of thickness $d_S$ separated from a normal metal $N$ of thickness $d_N$ by an insulating layer. Right: depiction of the processes leading to trapping: tunneling from $S$ to $N$ with rate $\Gamma_\text{tr}$ and from $N$ to $S$ with rate $\Gamma_\text{esc}(\epsilon)$, and relaxation in $N$ with rate $\Gamma_r$.} \label{fig:simple}
\end{figure}

Let us consider a quasiparticle trap made of a normal ($N$) metal covering part of a superconducting ($S$) qubit. The contact between the two superconductor and the normal trap is provided by an insulating ($I$) layer characterized by a small electron transmission coefficient.  In order to relate the quasiparticle tunneling rate to the conductance of the contact, we use the tunneling Hamiltonian formalism applied to a model $N$-$I$-$S$ system, see Fig.~\ref{fig:simple},
\bea
H & = & H_\qp + H_N + H_T\,, \\
H_\qp & = & \sum_{n\sigma} \epsilon_n \gamma^\dagger_{n\sigma} \gamma_{n\sigma} \,,\\
H_N & = & \sum_{m\sigma} \xi_m c^\dagger_{m\sigma} c_{m\sigma}\,, \\ \label{eq:H_T}
H_T & = & \frac{\widetilde{t}}{\sqrt{\Omega_N\Omega_S}} \sum_{m,n,\sigma} \left(c^\dagger_{m\sigma} d_{n\sigma} + d^\dagger_{n\sigma} c_{m\sigma} \right)\,.
\eea
We denote with $\Omega_{N,S}=A\times d_{N,S}$ the volumes of the $N$ and $S$ layers, respectively ($A$ is the area of interface, and $d_{N,S}$ are the layers thicknesses); $c^\dagger_{m\sigma}$ and $d^\dagger_{n\sigma}$ are the creation operators for electrons in the normal metal (energy $\xi _m$ and spin $\sigma$) and superconductor. The electron operators in the superconductor are related by Bogoliubov's transformation to the quasiparticle annihilation (creation) operators $\gamma^{(\dagger)}_{n\sigma}$,
\bea
d_{n\uparrow} & = & u_n \gamma_{n\uparrow} + v_n \gamma^\dagger_{n\downarrow} \\
d^\dagger_{n\downarrow} & = & -v_n \gamma_{n\uparrow} + u_n \gamma^\dagger_{n\downarrow} \\
u_n^2 &=& 1- v_n^2 = \frac12 \left(1 + \frac{\xi_n}{\epsilon_n}\right)\,.
\eea
Here $\epsilon_n = \sqrt{\xi_n^2+\Delta^2}$ is the energy of a quasiparticle, and $\xi_n$ is the energy of electron  in the normal state of the superconductor. The tunneling constant $\widetilde{t}$ can be related, by Fermi's golden rule, to the resistance $R_T$ of the contact,
\begin{equation}
\frac{R_q}{2\pi R_T} = 4\pi \left|\widetilde{t}\right|^2 \nu_{S0} \nu_{N0}\,, \quad R_q = \frac{2\pi\hbar}{e^2}\, ,
\label{eq:RT}
\end{equation}
where $\nu_{N0}$ and $\nu_{S0}$ are the densities of states in the normal metal and in the (normal state of the) superconductor, respectively. The tunnel conductance, $1/R_T$, is proportional to the area $A$ of the junction; the intensive quantity characterizing the insulating layer is its conductance per unit area, $1/R_TA$.

We may use Fermi's golden rule to evaluate also the rates of tunneling-induced change of the occupation factors of electrons, $f(\xi_m)=\sum_\sigma \langle c_{m\sigma}^\dagger c_{m\sigma}\rangle$, and quasiparticles, $f_\text{qp}(\epsilon_n)=\sum_\sigma \langle \gamma_{n\sigma}^\dagger \gamma_{n\sigma}\rangle$.
We can distinguish two processes. Quasiparticles tunnel from the superconductor into the normal metal with rate
$\Gamma_{\text{tr}} = 2\pi\left|\widetilde{t}\right|^{2}\nu_{N0}/\Omega_{S}$.
The transition rate is proportional to the density of the final states involved in the transition, therefore the quasiparticle trapping rate does not have a pronounced energy dependence. The complementary process of a non-equilibrium electron escape into the superconductor, however, does display a strong energy dependence associated with the BCS singularity in the density of final states, $\Gamma_{\text{esc}}\left(\epsilon\right) = 2\pi\left|\widetilde{t}\right|^{2}\nu_{S0} \nu_S(\epsilon)/\Omega_{N}$; here
\be\label{BCSdos}
\nu_S(\epsilon)=\frac{\epsilon}{\sqrt{\epsilon^2-\Delta^2}}
\ee
is the normalized BCS density of states.

One can see from Eq.~(\ref{eq:RT}) that the rates $\Gamma_{\text{tr}}$ and $\Gamma_{\text{esc}}\left(\epsilon\right)$ are independent of the area $A$ at fixed conductance per unit area of the insulating layer. We may express the rates as
\begin{equation}\label{eq_gamma_esc}
\Gamma_\text{tr}=\widetilde{\gamma}_\text{tr}/{d_S}\,,\quad
\Gamma_\text{esc}(\epsilon)=\widetilde{\gamma}_\text{esc}(\epsilon)/{d_N}
\end{equation}
in terms of quantities independent of geometry,  $\widetilde{\gamma}_\text{tr}$ and $\widetilde{\gamma}_\text{esc}$,
\be\label{Gtrdef}
\widetilde{\gamma}_\text{tr} = \frac{R_q}{4\pi (R_T A)\nu_{S0}}\,,\quad
\widetilde{\gamma}_\text{esc} = \frac{R_q\nu_{S}(\epsilon)}{4\pi (R_T A)\nu_{N0}}\,.
\ee
with $(R_T A)$ being the contact resistance times the area of the contact. This product, with units of $\Omega\cdot$cm$^2$, is independent of $A$, being inversely proportional to the transmission coefficient through the insulating barrier.

The above formulas enable us to estimate the trapping and escape rates for an aluminum-copper interface for a typical experimental setup (cf. Sec.~\ref{sec:exp}): aluminum has a density of states $\nu_{S0} = 0.73\times10^{47}/$Jm$^3$ \cite{aldos} and
a direct measurement of the contact resistance yields $(R_TA)\sim430\,\Omega \mu \text{m}^2$ (this corresponds to the transmission coefficient of order $10^{-5}$). Taking $d_S \sim 80\,$nm we find, using Eqs.~(\ref{eq_gamma_esc}) and (\ref{Gtrdef}), $\Gamma_\text{tr} \sim 8\cdot10^{6}\,{\rm s}^{-1}$.
The escape rate saturates at an energy-independent value, $\Gamma_\text{esc}(\epsilon)\to \Gamma_\text{esc}$ at energies $\epsilon\gg\Delta$. Since in $d_S\approx d_N$ and $\nu_{S0}\approx\nu_{N0}$ in a typical experiment, one has $\Gamma_\text{esc}\approx\Gamma_\text{tr}$.

In writing the rate equations for the energy distribution functions of electrons and quasiparticles, we assume the continuum limit for energies $\xi_m$ and $\epsilon_n$. It is convenient to define the probability density to find an electron (quasiparticle) in the normal metal (superconductor) with energy $\epsilon\geq\Delta$ as
\begin{align}
p_N(\epsilon)=&\frac{\nu_{N0}{\Omega_N}}{\nu_{S0}{\Omega_S}}f(\epsilon) \label{pn_def}\\
p_S(\epsilon)=&\nu_S(\epsilon) f_\text{qp}(\epsilon)\, . \label{ps_def}
\end{align}
Without loss of generality, we normalize the probability with respect to $\nu_{S0}\Omega_S$.
Note that eventually, the experimentally accessible quantity is the normalized quasiparticle density, which can be derived from $p_S$ as
\begin{align}
x_\text{qp}=&\frac{2}{\Delta}\int_\Delta^\infty d\epsilon\,p_S(\epsilon)\ .
\end{align}
In the absence of spatial dispersion of the distribution functions, the rate equations read (see Appendix~\ref{app:backflow_rate_eq})
\begin{align}\label{eq:dot_p_N}
\dot{p}_{N}\left(\epsilon\right) = & \Gamma_{\text{tr}}p_{S}\left(\epsilon\right)-\Gamma_{\text{esc}}\left(\epsilon\right)p_{N}\left(\epsilon\right)-\Gamma_{\text{r}}p_{N}\left(\epsilon\right)\, ,\\
\label{eq:dot_p_S} \dot{p}_{S}\left(\epsilon\right) = & \Gamma_{\text{esc}}\left(\epsilon\right)p_{N}\left(\epsilon\right)-\Gamma_{\text{tr}}p_{S}\left(\epsilon\right)\,.
\end{align}
The terms proportional to $\Gamma_\text{tr}$ describe trapping of quasiparticle excitations in the normal metal, and those proportional to $\Gamma_\text{esc}(\epsilon)$ the possible escape of electron excitations back to the superconductor; these events take place with rates described by Eqs.~(\ref{eq_gamma_esc})-(\ref{Gtrdef}).

Since the tunneling process is elastic, excitations appear in the normal metal at energies close to the gap $\Delta$. At low temperature $T\ll\Delta$, there are many unoccupied states below $\Delta$ in the normal metal, into which the excitations can decay. These inelastic processes are mediated by electron-electron and electron-phonon interactions and lead to relaxation, which we capture in \eref{eq:dot_p_N} with the phenomenological rate $\Gamma_\text{r}$. All the processes included in the rate equations \rref{eq:dot_p_N}-\rref{eq:dot_p_S} are represented in the right panel of Fig.~\ref{fig:simple}.

If the relaxation is immediate, the quasiparticles get trapped in the normal metal with rate $\Gamma_\text{tr}$. However, the relaxation rate $\Gamma_\text{r}$ due to electron-electron and electron-phonon interactions in the normal metal is of course finite. It has been estimated in the supplementary to \ocite{wang} to be $\Gamma_r\sim 10^7\,{\rm s}^{-1}$; the measurements reported in Ref.~\cite{anthore} lead to a relaxation rate for electron-phonon interaction of the same order of magnitude, while an estimate based on \ocite{BH-thesis} yields the faster relaxation rate $\Gamma_r\sim 10^8\,{\rm s}^{-1}$. In all cases, relaxation cannot be assumed immediate in comparison with the trapping and escape rates estimated above, especially taking into account that the escape rate quickly increases for energies approaching the gap due to the divergent BCS density of states in \eref{BCSdos}. In fact, for some energy interval close to the gap, the escape rate dominates the quasiparticle dynamics, such that the excitations do not have enough time to relax. Therefore, we cannot in general neglect the backflow of excitations from the normal trap to the superconductor.

The backflow may result in an effective rate which is slower than $\Gamma_\text{tr}$.
Assuming a steady-state distribution of non-equilibrium electrons in the normal layer, we set $\dot{p}_N=0$ in Eq.~\eqref{eq:dot_p_N} and solve for $p_N$ in terms of $p_S$ (see also App.~\ref{app:diffusion_eq}). Substituting the solution into \eref{eq:dot_p_S} and integrating over energy, we arrive at
\begin{equation}\label{eq:0dxpq}
\dot{x}_\text{qp}=-\Gamma_\text{eff}x_\text{qp}\, ,
\end{equation}
with the effective trapping rate defined by
\begin{equation}\label{eq:Geffdef}
\Gamma_\text{eff}=\frac{1}{\int_\Delta^\infty d\epsilon\,p_S(\epsilon)}\int_\Delta^\infty d\epsilon\,\frac{\Gamma_\text{tr}\Gamma_\text{r}}{\Gamma_\text{esc}(\epsilon)+\Gamma_\text{r}}p_S(\epsilon)\,.
\end{equation}
It is clear that $\Gamma_\text{eff}$ is suppressed to a level below $\Gamma_\text{tr}$. The level of suppression depends on the typical width of the quasiparticle distribution function in energy space. Assuming $p_S(\epsilon)$ is characterized by an effective temperature, $T\ll\Delta$, we find that the trapping is not suppressed,  $\Gamma_\text{eff}\approx\Gamma_\text{tr}$, only if the energy relaxation is fast enough ($\Gamma_\text{r}\gg (\Delta/T)^{1/2}\Gamma_\text{esc}$); in this case excitations in the normal metal quickly relax to energies below the gap and cannot return into the superconductor. In the opposite case ($\Gamma_\text{r}\lesssim (\Delta/T)^{1/2}\Gamma_\text{esc}$), the effective rate becomes $T$-dependent and suppressed below the nominal trapping rate, $\Gamma_\text{eff}\approx (2T/\pi\Delta)^{1/2}\Gamma_\text{tr}\Gamma_\text{r}/\Gamma_\text{esc}$.
Note that in the slow relaxation regime the effective trapping rate $\Gamma_\text{eff}$ is independent of the tunneling probability between superconductor and normal metal, the limiting value of $\Gamma_\text{eff}$ being proportional to the relaxation rate.

The quasi-static approximation ($\dot{p}_N=0$) we used above becomes justified once we move from the model system of Fig.~\ref{fig:simple} to a more realistic geometry of a long superconducting strip in contact with a metallic trap, see Fig.~\ref{fig:sketch}a. In that geometry, the time variation of the quasiparticle distribution function $p_S$ is controlled by the diffusion time in the strip, which is typically substantially longer than $1/\Gamma_{\rm r}$.
The generalization of the rate equations \rref{eq:dot_p_N} and \rref{eq:dot_p_S} to include diffusion is performed in Appendix~\ref{app:diffusion_eq}. In addition to diffusion, other processes such as quasiparticle recombination, generation, and trapping in the bulk must be generally taken into account. For sufficiently thin normal and superconducting layers, we find a generalized diffusion equation for the quasiparticle density $x_\qp$,
\begin{align}
\label{eq:diff_eq_qp}
\dot{x}_\text{qp}=&D_\text{qp}\nabla^2x_\text{qp}-a(x,y)\Gamma_\text{eff}x_\text{qp}\\\nonumber
&-r x_\text{qp}^2 - s_\text{b} x_\text{qp} + g \, ,
\end{align}
where $x_\text{qp}(x,y)$ depends only on coordinates in the plane of the superconducting strip (and is assumed constant across its thickness) and the area function $a(x,y)$ equals 1 for $x$ and $y$ where the trap and the superconductor are in contact, and 0 elsewhere, see Fig.~\ref{fig:sketch}(a).

The diffusion constant $D_\qp$ in \eref{eq:diff_eq_qp} is proportional to the normal-state diffusion constant for the electrons in the superconductor -- the proportionality coefficient can in principle be calculated from the detailed information on the energy distribution of quasiparticles that we discard in using the phenomenological \eref{eq:diff_eq_qp}. The recombination term $r x_\qp^2$ accounts for processes in which two quasiparticles recombine into a Cooper pair \cite{rt}, again neglecting the details of the quasiparticle distribution. The relationship between recombination time, quasiparticle energy, and electron-phonon interaction strength can be found in \ocite{scalapino}. Moreover, there is a background trapping term $s_\text{b}x_\text{qp}$ that describes any process that can localize a quasiparticle and hence remove its contribution to the bulk density $x_\qp$. Trapping by vortices is an example of such a process, recently characterized in \ocite{wang}. The generation rate $g$ describes pair-breaking processes, both thermal and non-thermal; at low temperatures, non-equilibrium processes of unknown origin lead to a quasiparticle density orders of magnitude larger than the thermal equilibrium one \cite{riste,klapwijk}.

In what follows we will neglect both background trapping and recombination: according to the measurements in \ocite{wang} we expect $s_\text{b}<0.2\cdot10^{3}\,{\rm s}^{-1}$ as well as $r x_\text{qp}<1.25\cdot10^{3}\,{\rm s}^{-1}$ (having assumed $x_\text{qp}<10^{-4}$). Both processes are orders of magnitudes slower than the effective trapping rate $\Gamma_\text{eff}$, even when the latter is highly reduced by backflow. Indeed, even for a low effective temperature $T=10\,\text{mK}$, using $\Gamma_\text{r}\sim 10^7\,{\rm s}^{-1}$ and $\Delta/h = 44\,$GHz for aluminum, we find $\Gamma_\text{eff}\sim 0.55\cdot10^6\,{\rm s}^{-1}$.
Finally, we assume a long wire geometry, where the dimensions of the system in the $x$ and $z$ directions are sufficiently small such that the superconductor can be treated as (quasi)one-dimensional, and we consider traps that are small (in a sense to be specified below), so that they are effectively zero-dimensional.
In this case, from \eref{eq:diff_eq_qp} we obtain
\begin{equation}\label{eq:diff_eq_1D}
\dot{x}_\text{qp}=D_\text{qp}\partial_y^2x_\text{qp}-\gamma\delta(y-l)x_\text{qp}+ g\,,
\end{equation}
where the trap is at position $y=l$ and $\gamma=\Gamma_\text{eff}\,d$, with $d$ the length of the trap in $y$ direction. To estimate when the trap is sufficiently small, we note that the trapping length
\be\label{ltrdef}
\lambda_{\rm tr} \equiv \sqrt{D_\qp/ \Gamma_{\rm eff}}
\ee
gives the scale over which the density decays due to trapping, so the smallness condition is $d \ll \lambda_{\rm tr}$.
In the next section we study the dynamics of the quasiparticle density by solving \eref{eq:diff_eq_1D} in various regimes.

\section{Quasiparticle dynamics during injection and trapping}
\label{sec:dyn}

\begin{figure}
\begin{center}
\includegraphics[width=0.4\textwidth]{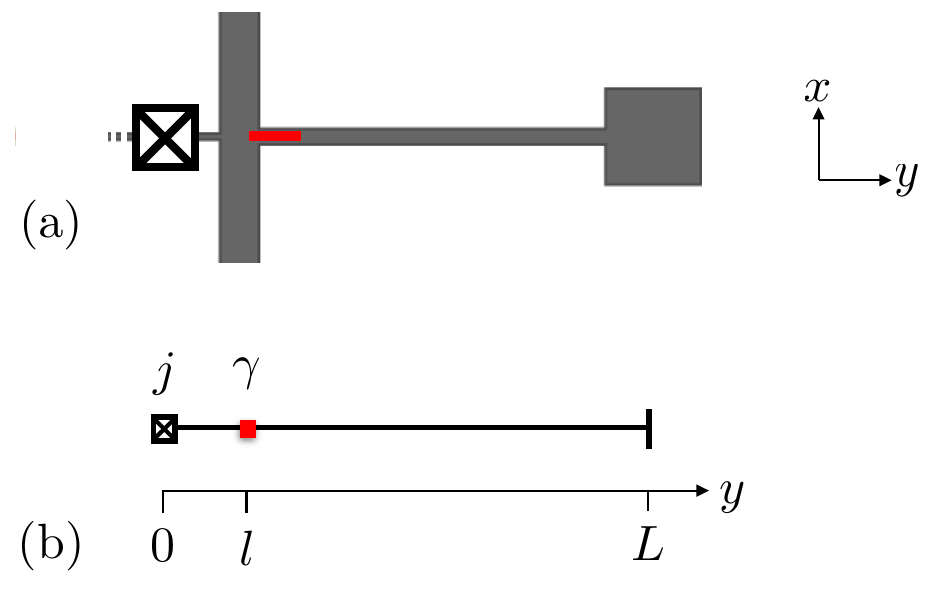}
\end{center}
\caption{a) Figure of a realistic transmon qubit device close to the proportions of experiment. The Josephson junction is indicated with the crossed box, in grey is the superconductor, and in red the normal metal trap. Shown is half the qubit (the dashed lines indicate that the superconducting structure including trap is mirrored on the left hand side of the junction).
b) Simplified model of a 1D superconducting strip with small trap, described by Eq.~\eqref{eq:diff_eq}. \label{fig:sketch}}
\end{figure}

In this section we compute the dynamics of the quasiparticle density in a simple geometry depicted in Fig.~\ref{fig:sketch}(b). It models a transmon qubit in Fig.~\ref{fig:sketch}(a) by neglecting for simplicity both the gap capacitor near the Josephson junction and the square pad at the opposite end of the long wire.
Note that because of the spatial symmetry, it is sufficient to consider only half of the system, $0\leq y\leq L$.
After separating out the steady-state background density due to the finite generation rate $g$, the equation controlling the evolution of the excess density of quasiparticles takes the form
\begin{equation}\label{eq:diff_eq}\begin{split}
\frac{\partial x_{\qp}\left(y,t\right)}{\partial t}=\left[D_{\qp}\frac{\partial^{2}}{\partial y^{2}}-\gamma\delta\left(y-l\right)\right]x_{\qp}\left(y,t\right) \\ + j \delta\left(y-0^{+}\right)\theta\left(-t\right)\theta\left(t+t_{\text{inj}}\right).
\end{split}\end{equation}
This diffusion equation is supplemented by the boundary conditions $\partial_{y}x_{\qp}\left(L,t\right)=0$ and $\partial_{y}x_{\qp}\left(0,t\right)=0$. The former condition ensures that no quasiparticles leave the device (hard wall condition), while the latter reflects the spatial symmetry of the system.

In the experiments, quasiparticles are generated at the Josephson junction when injecting a high-power microwave pulse into the cavity hosting the qubit~\cite{wang}, resulting in a time-dependent source of quasiparticles localized at $y=0$. In \eref{eq:diff_eq}, this source is modeled by a term with a generation rate proportional to $j$ active over the time interval $-t_{\text{inj}}<t<0$.
Clearly, there are two stages of time evolution: first, during the injection process, when the source term is switched on, the quasiparticle density will
start to rise and distribute across the wire.
Once the source term is switched off, the presence of the normal-metal trap ensures the decay of the excess density back to zero. In the following, we provide analytical results for the time-dependent dynamics of the quasiparticle density, where we focus predominantly on the experimentally accessible~\cite{wang} density at the junction, $y=0$.

The time-dependent diffusion equation \rref{eq:diff_eq} can be solved via a
decomposition in the modes $e^{\lambda_k t}n_{k}\left(y\right)$ of the homogeneous equation (i.e., \eref{eq:diff_eq} without the source term), with $\lambda_{k}$ being the eigenvalue and $n_k$ satisfying equation
\begin{equation}
\lambda_{k}n_{k}\left(y\right)=\left[D_{\text{qp}}\frac{\partial^{2}}{\partial y^{2}}-\gamma\delta\left(y-l\right)\right]n_{k}\left(y\right).\label{eq:eigenmodes}
\end{equation}
For a strip of finite length $L$, the eigenvalues are discrete and the eigenmodes form an orthonormal basis,
\begin{equation}
\int_{0}^{L}\frac{dy}{L}n_{k}\left(y\right)n_{k'}\left(y\right)=\delta_{kk'}.\label{eq:completeness}
\end{equation}
In presence of the trap at $y=l$, the eigenmodes are defined piecewise as
\begin{equation}
n_{k}\left(y\right)=\frac{1}{\sqrt{N_{k}}}\left\{ \begin{array}{cc}
\cos\left(ky\right)&\quad y<l\\
a_{k}\cos\left(ky\right)+b_{k}\sin\left(ky\right)&\quad y>l,
\end{array}\right.\label{eq:eigenfunctions}
\end{equation}
with the normalization constant $N_{k}$ (which will be provided explicitly
later in some limiting cases) and the coefficients
\begin{eqnarray}
a_{k} & = & 1-\frac{\gamma}{D_\qp k}\cos\left(kl\right)\sin\left(kl\right)\nonumber \\
b_{k} & = & \frac{\gamma}{D_\qp k}\cos^{2}\left(kl\right).\label{eq:coefficients}
\end{eqnarray}
The  eigenvalue corresponding to eigenmode $k$ is $\lambda_{k}=-D_\qp k^{2}$. The boundary condition at $y=0$ is satisfied by \eref{eq:eigenfunctions}, while the one at $y=L$ gives the equation
\begin{equation}
\cot\left(kL\right)=\frac{1-\frac{\gamma}{D_\qp k}\cos\left(kl\right)\sin\left(kl\right)}{\frac{\gamma}{D_\qp k}\cos^{2}\left(kl\right)}.\label{eq:bc_hard_wall}
\end{equation}
which fixes the wave vector $k$ to discrete values.

In terms of the eigenbasis introduced above, by solving \eref{eq:diff_eq} we find that the excess quasiparticle density immediately
after the injection, at time $t=0$, is given by
\begin{equation}
x_{\qp}\left(y,0\right)=\sum_{k}c_{k}\frac{e^{\lambda_{k}t_{\text{inj}}}-1}{\lambda_{k}}n_{k}\left(y\right)\label{eq:x_tinj}
\end{equation}
with
\begin{equation}
c_{k}=j \int_{0}^{L}\frac{dy}{L}n_{k}\left(y\right)\delta\left(y-0^{+}\right)=\frac{j}{L}n_{k}\left(0\right).\label{eq:alpha_k}
\end{equation}
where we assumed that at times $t<-t_{\text{inj}}$, there were no excess
quasiparticles in the system. Once the injection stage is finished, the subsequent trapping of the quasiparticles controls the evolution of their density,
\begin{equation}
x_{\qp}\left(y,t\right)
 =\frac{j}{L}\sum_{k}n_{k}\left(0\right)e^{-D_\qp k^{2}t}\frac{1-e^{-D_\qp k^{2}t_{\text{inj}}}}{D_\qp k^{2}}n_{k}\left(y\right). \label{eq:x_t}
\end{equation}
The here derived expressions for $x_{\qp}\left(y,t\right)$ are general and do not rely on any further simplifying assumption. Next, we consider in more detail several limiting cases.

\subsection{The long-strip limit}
\label{sec:semiinfw}

If both the injection time $t_{\rm inj}$ and the time $t$ after injection are short compared to the diffusion time scale $\sim L^2/D_{\qp}$, the generated quasiparticles do not reach the far end of the strip, and we may take the limit $L\to\infty$. In this limit, all values of $k$ are allowed and sums over $k$ are replaced by an integral, $\frac{1}{L}\sum_k\rightarrow \int \frac{dk}{2\pi}$. Moreover, when letting $L\rightarrow\infty$ while keeping the distance $l$ between trap and junction finite, the normalization constant $N_k$ is dominated by the part of the mode with $y>l$, so that $N_k \simeq (a_k^2+b_k^2)/2$. Clearly, a single trap suppresses the excess quasiparticle density at the junction best if the distance $l$ is short. For simplicity, from now on we assume  $l\rightarrow 0^+$. That leaves us with only one characteristic time scale, the saturation time
\be\label{eq:tsatdef}
t_\text{sat}=D_\qp/\gamma^2\, .
\ee
It gives the time scale over which the density near the junction approaches its steady-state value $x_0 = j/\gamma$, prescribed by the balance between generation and trapping, during the injection process.
Indeed, after time $\tau$ from the start of the injection, quasiparticles have spread over a distance $\sim \sqrt{D_{\rm qp}\tau}$ and the diffusive current at that time can be estimated as $D_{\rm qp}\partial_y x_{\qp}(0)\sim D_{\rm qp} x_{\qp}(0)/\sqrt{D_{\rm qp}\tau}$. For $\tau=t_{\rm sat}$ the diffusive current is therefore of the order of the trapping current $\gamma x_\qp(0)$; as quasiparticles spread further out, the diffusive current will decrease, indicating that indeed a steady-state is (asymptotically) reached.
It is important to note that the total number of quasiparticles in the device keeps growing for the entire duration of injection, despite the saturation of $x_{\qp}(0)$ at $\tau\sim t_{\rm sat}$.

The evolution in the relaxation stage, $t>0$, depends on the ratio $t_\text{sat}/t_{\rm inj}$. A straightforward use of Eq.~\eqref{eq:x_t} yields for the quasiparticle density close to the trap, $y\rightarrow0$, in the long-time limit $t\gg t_\text{sat}$
\begin{equation}
\label{eq:x_ty_inf}
x_\text{qp}(0,t) \approx \frac{x_{0}}{\sqrt{\pi}}\left(\sqrt{\frac{t_{\text{sat}}}{t}}-\sqrt{\frac{t_{\text{sat}}}{t+t_{\text{inj}}}}\right)\,.
\end{equation}
This asymptote is valid for any value of $t_{\rm sat}/t_{\rm inj}$. If $t_{\rm inj}\gg t_{\rm sat}$, one may distinguish between an intermediate asymptotic behavior, $x_{\rm qp}(0,t)\propto 1/t^{-1/2}$, valid at times $t_\text{sat}\ll t\ll t_{\rm inj}$, and a long-time asymptote, $x_{\rm qp}(0,t)\propto t^{-3/2}$, at $t\gg t_{\rm inj}$. Only the latter behavior is present for short injection times $t_{\rm inj} \lesssim t_{\rm sat}$.

\subsection{The effect of finite diffusion time}\label{sec:finite_diff_time}

We now turn to the case of a finite-length strip, so that the diffusion time across the whole device,
\be
t_L=4L^2/(\pi^2D_{\qp}) \, ,
\label{tL}
\ee
provides yet another scale for the relaxation dynamics of $x_{\qp}$.
The comparison of the two time scales, $t_L$ and $t_{\rm sat}$, allows us to introduce the notion of a weak versus a strong trap. A weak trap corresponds to $t_\text{sat}\gg t_L$. The diffusion through the device occurs much faster than the local saturation at the trap, and consequently, the quasiparticle distribution is almost homogeneous throughout the device. A strong trap, $t_\text{sat}\ll t_L$, leads to a highly-inhomogeneous spatial distribution of the quasiparticle density.
Recalling that $\gamma=\Gamma_\text{eff}d$, this distinction can also be expressed in terms of a comparison of the trap length $d$ with the length scale
\be
l_0\equiv \frac{\pi}{2} \frac{D_{\qp}}{L\Gamma_{\rm eff}}=\frac{\pi}{2}\frac{\lambda_\text{tr}^2}{L}\, ,
\label{l0}
\ee
with $\lambda_{\rm tr}$ of \eref{ltrdef}; a weak (strong) trap is characterized by $d\ll l_0$ ($d\gg l_0$). Note that if $\lambda_\text{tr}\ll L$, $l_0$ is much smaller than $\lambda_\text{tr}$, so the crossover between the two limits occurs while the trap length remains short, $d\ll\lambda_{\rm tr}$, and we can still use Eq.~({\ref{eq:diff_eq}).

For a weak trap, $d\ll l_0$, we may neglect the $y$-dependence of $x_{\qp}(y,t)$ in \eref{eq:diff_eq}, and integrating it over $y$ we find
\begin{equation}
\label{short}
x_\text{qp}(y,t)\approx x_{0}\left(1-e^{-t_{\text{inj}/\tau_w}}\right)e^{-t/\tau_w},
\end{equation}
where
\be\label{tauw}
\frac{1}{\tau_w} = \frac{d}{L}\Gamma_{\rm eff} \, .
\ee
As long as $x_\text{qp}$ can be considered $y$-independent, the expression \rref{tauw} for the density decay rate may be easily generalized: the ratio $d/L$ in the right hand side
should be replaced by  $A_\text{tr}/A_\text{dev}$, where $A_\text{tr}$ is the total area of the trap and $A_\text{dev}$ is the area of the entire device. Importantly, the decay rate here depends merely on the ratio of the total areas, whereas details of the geometry of the trap and device are unimportant.

In the opposite case of a strong trap, $d\gg l_0$, the approximation of a constant $x_\text{qp}(x,y)$ is no longer valid, and the decay rate will depend on the details of the trap geometry and placement. For simplicity, we concentrate again on the strip geometry.
To obtain the eigenmodes, one may replace the right hand side of \esref{eq:bc_hard_wall} by zero. Therefore, $k$ is
simply given by $k=\frac{\pi}{2L}p$, where $p$ is an odd integer (up to small corrections of order $l_0/d$ -- cf. \eref{eq:k_zero_numeric}). In contrast to the case of a weak trap, the relaxation is now limited by the diffusion time. From \eref{eq:x_t} we find the time-dependent quasiparticle density at the junction to be
\begin{equation}
x_{\qp}\left(0,t\right)\approx \frac{4}{\pi}x_{0}\sqrt{\frac{t_{\text{sat}}}{t_{L}}}\sum_{p}
\left[e^{-\frac{t}{t_{L}}p^{2}}-e^{-\frac{t+t_{\text{inj}}}{t_{L}}p^{2}}\right],\label{eq:x_t_junction}
\end{equation}
with $x_{0}=j/\gamma$,  and the sum over the odd integer $p$.
For short times, $t\ll t_{L}$, the time evolution is insensitive to the
boundary condition at $x=L$, and indeed we recover the results given in Sec.~\ref{sec:semiinfw}. (Note that of course, being able to observe the transition from a $t^{-1/2}$ to a $t^{-3/2}$ power law decay is contingent upon $t_\text{inj}$ being much smaller than $t_L$.)
For times exceeding the diffusion time, $t \gtrsim t_{L}$, the time-evolution is dominated by the single exponential of the slowest mode, and we can write
\be
x_\qp \left(0,t\right)\approx \frac{4}{\pi}x_{0}\sqrt{\frac{t_{\text{sat}}}{t_{L}}} \left(1-e^{-t_\text{inj}/\tau_{w}}\right) \, e^{-t/\tau_{w}}
\ee
where the decay time constant is now determined by the diffusion time (\ref{tL}), $\tau_w=t_L$~\cite{note_tL}.

Concentrating on the long-time evolution, we can more generally relate the time constant $\tau_{w}$ to the wave number of the slowest mode. Thus, we are able to investigate the full crossover in $\tau_{w}$ from weak to strong trap as a function of $d/l_0$. Setting $l\to 0$ in Eq.~\eqref{eq:bc_hard_wall}, we may re-write it as
\begin{equation}\label{eq:k_zero_numeric}
\cot{\left(\frac{\pi}{2}\widetilde{k}\right)}=\frac{l_0}{d}\widetilde{k}\,,\,\,\widetilde{k}=\frac{2}{\pi}kL\,.
\end{equation}
The time constant can be expressed in terms of the smallest positive solution $\widetilde{k}_0$ of Eq.~\eqref{eq:k_zero_numeric} as $\tau_{w}=t_L/\widetilde{k}_0^2$. Therefore, the ratio $t_L/\tau_w$ is a function of a single variable, $d/l_0$. The full crossover function between the linear dependence at small $d/l_0$ and saturation at $d/l_0\gg 1$ can be found by solving Eq.~\eqref{eq:k_zero_numeric} numerically.
In Fig.~\ref{fig:trapping_rate_fit}, we show $t_L/\tau_w$ as a function of $d/l_0$, together with experimental data that we discuss in the next section. The introduction of scaled variables $t_L/\tau_w$ and $d/l_0$ allows us to compare the trapping for a number of devices and for a set of different temperatures.

\section{Experimental Data}
\label{sec:exp}

In this section we compare the model developed in the previous sections with experiments measuring the dynamics of injected quasiparticles in 3D transmon qubits~\cite{3dtr}. The qubit, similar to the device sketched in Fig. \ref{fig:sketch}, consists of a single Al/AlO$_x$/Al Josephson junction shunted by a coplanar gap capacitor, with long ($\sim$1 mm), narrow antenna leads which connect to a pair of small ($80\,\times\, 80\, \mu$m$^2$) pads, see Fig. \ref{fig:device_picture}. One or two chips containing qubits are mounted in a superconducting aluminum rectangular waveguide cavity. All measurements are performed in an Oxford cryogen-free dilution refrigerator, with magnetic field shielding, infrared shielding and filtering described in Ref.~\cite{sears}.

After fabrication of the qubits, normal-metal traps are patterned via optical lithography, which gives control of trap location and size to better than $1~\mu$m. The heavily oxidized aluminum surface of the qubit is treated with an ion etch, and 100 nm of copper is deposited in a liftoff process thereafter. Through independent DC measurements, we find the Al-Cu interface resistance to be between 200 and 430$\,\Omega\cdot \mu$m$^2$. As shown in Fig.~\ref{fig:device_picture}c, one edge of the trap is located a short, fixed distance ($\sim 35\,\mu$m) away from the junction. The trap has a width of $8\,\mu m$, and it is placed symmetrically on the $12\,\mu m$ wide lead. For this study, we focus on devices in which the trap length, $d$, along the lead is varied from 20 to $400~\mu$m.

\begin{figure}[tb!]
\includegraphics[scale = 0.55]{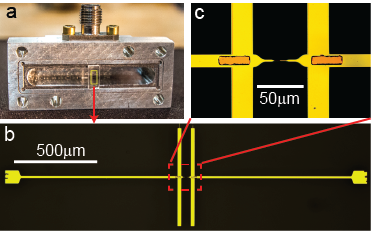}
\caption{a) Photograph of a 3D aluminium cavity loaded with a transmon qubit. b) Optical image of an example of the devices used for this study. c) Zoomed-in image of the Cu-trap deposited near the junction}\label{fig:device_picture}
\end{figure}

The QP dynamics of these devices is studied using the  contactless, \textit{in-situ} method described in Ref.~\cite{wang}, where QPs are introduced into the qubit by applying a large microwave tone at the bare cavity resonance. This injection pulse creates a voltage across the Josephson junction greater than $2\Delta$, generating many ($\gtrsim 10^5$ per $\mu$s) quasiparticles near the junction. The subsequent decay of $x_{\mathrm{qp}}$ is probed by monitoring the recovery of the qubit relaxation time $T_1$ measured as a function of time after the injection, in light of the simple relation
\be
\Gamma (t) = 1/T_{1} (t) = C x_{\mathrm{qp}}(0, t) + \Gamma_{0}\, ,
\ee
where $\Gamma_0$ is the steady-state relaxation rate of the qubit, which includes the effects of residual quasiparticle population and other relaxation mechanisms such as dielectric losses, and $C$ is a known proportionality constant~\cite{prb1} (whose value we do not need here).
In other words, we exploit the fact that the time-dependent part of the qubit decay rate $\Gamma$ is directly proportional to the excess quasiparticle density at the junction, $y=0$.

\begin{figure}[bt!]
\centering
\includegraphics[scale = 0.49]{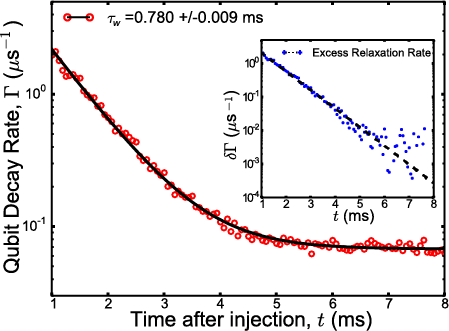}
\caption{Qubit energy relaxation rate $\Gamma$ after quasiparticle injection. The solid line is a fit to the data by a single exponential with time constant $\tau_w$, see \eref{Gt}. The inset shows the relaxation rate after subtracting a constant background in logarithmic scale, displaying good agreement with the predicted functional form.}
\label{fig:qp_decay}
\end{figure}

Figure~\ref{fig:qp_decay} shows a typical measurement of the qubit decay rate in a device with a small normal-metal trap.
The decay time constant $\tau_w$ is estimated by fitting the data with a single exponential of the form
\be\label{Gt}
\Gamma (t) = A e^{-t/\tau_w} + \Gamma_0\ .
\ee
As discussed in Sec.~\ref{sec:finite_diff_time}, we are considering only the slowest decay mode of $x_{\qp}$, so we fit the data to the above expression at long times $t\gtrsim t_L$ [with $t_L$ of \eref{tL}], where we find good agreement between the data and the predicted single-exponential decay.

Repeating the measurement for several trap lengths $d$, we find that the experimental decay rate $1/\tau_w$ varies with the length of the trap in qualitative agreement with the rate calculated by solving Eq.~\eqref{eq:k_zero_numeric}, see Fig. \ref{fig:trapping_rate_fit}.
Indeed, for short traps we approximately find the linear dependence of $1/\tau_w$ on the trap length predicted by \eref{tauw},
while for longer traps the rate saturates to the the diffusion limit, $1/\tau_w\approx 1/t_L$.
To scale the experimental data so that they can be compared to the theoretical expectation, we use $l_0$ and $t_L$ as fitting parameters, and allow them to be different for data taken at different fridge temperatures $T_\text{fr}$, thus assuming that both $D_\text{qp}$ as well as $\Gamma_\text{eff}$ depend on $T_\text{fr}$. The fitting parameters are $l_0=41.2\pm 17.1\,\mu\text{m}$ and $t_L=184\pm 29\,\mu\text{s}$ for $T_\text{fr}=13\,$mK and $l_0=45.8\pm 16.7\,\mu\text{m}$ and $t_L=125\pm 20\,\mu\text{s}$ for $T_\text{fr}=50\,\text{mK}$ \cite{notefit}. Note that the relative change in $l_0$ is smaller than that in $t_L$ and that this is in qualitative agreement with theoretical expectations: since $l_0$ is proportional to $D_\qp/\Gamma_\text{eff}$, the expected increases of both $D_\qp$ and $\Gamma_\text{eff}$ with effective temperature can partially compensate each other, while no such compensation is possible for $t_L \propto 1/D_\qp$.

\begin{figure}[b!]
\begin{center}
\includegraphics[scale=.43]{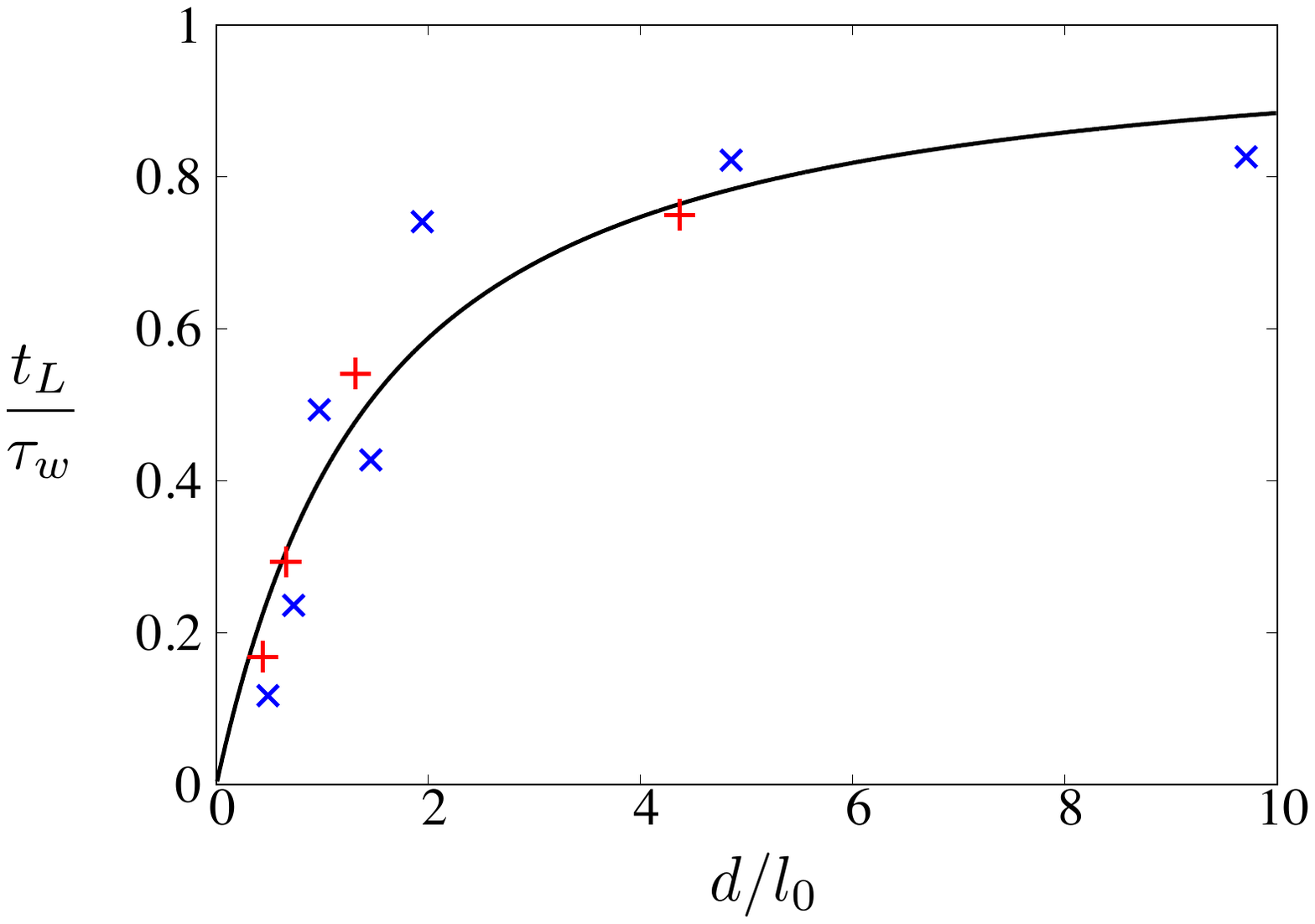}
\end{center}
\caption{Dimensionless density decay rate $1/\tau_w$ normalized by the diffusion time $t_L$, cf. Eq.~(\ref{tL}), as a function of trap length $d$ measured in units of $l_0$, see Eq.~(\ref{l0}) for the definition. The solid line is calculated by solving Eq.~\eqref{eq:k_zero_numeric} numerically. The experimental data are taken at two different fridge temperatures: the blue symbol ``x'' is used for $T_\text{fr}=13\,$mK and the red symbol ``+'' for $T_\text{fr}=50\,$mK. Note the transition from a linear dependence to the saturated diffusive limit at $d\sim l_0$.}
\label{fig:trapping_rate_fit}
\end{figure}

As discussed after \eref{tauw}, in the linear regime we can take into account the actual geometry of the transmon by modifying that expression for the decay rate, which becomes $1/\tau_w =\Gamma_\text{eff}A_\text{tr}/A_\text{dev}$. We use this formula to estimate $\Gamma_\text{eff}$ using the short-trap data and find $\Gamma_\text{eff}\approx 2.42\!\cdot\!10^{5}\,{\rm s}^{-1}$ for $T_\text{fr}=13\,\text{mK}$ (corresponding to the blue data points in Fig.~\ref{fig:trapping_rate_fit}) and $\Gamma_\text{eff}\approx 3.74\!\cdot\!10^{5}\,{\rm s}^{-1}$ for $T_\text{fr}=50\,\text{mK}$ (red points). These numbers are close to the order-of-magnitude estimate for $\Gamma_\text{eff}$ given at the end of Sec.~\ref{sec:diff}, where we assumed that the backflow of quasiparticles must be taken into account and strongly suppresses the effective trapping rate. In that Section we have also shown that $\Gamma_\text{eff}\sim\Gamma_\text{r}(T/\Delta)^{1/2}$, indicating that $\Gamma_\text{eff}$ should grow with temperature. While we observe an increase in the $\Gamma_\text{eff}$ extracted from the data with increasing fridge temperature, this increase is smaller than the factor of 2 expected from theory. This discrepancy is not surprising, since it is known that at low temperatures the quasiparticles are not in thermal equilibrium at the fridge temperature~\cite{riste}. Moreover, the injection pulse can cause additional heating in the qubit~\cite{vool}, further weakening the relationship between fridge temperature and quasiparticle effective temperature.

\section{Summary}
\label{sec:summ}

In this work we develop a basic model enabling us to predict the effect of a normal-metal trap on the dynamics of the nonequilibrium quasiparticles population in a superconducting qubit. The model accounts for the tunneling between the superconductor and the trap, as well as for the electron energy relaxation in the trap, see \eref{eq:Geffdef}. The surprising finding is that the effective trapping rate $\Gamma_\text{eff}$ is sensitive to the energy of the quasiparticles and is constrained by their backflow from the normal-metal trap on time scales shorter than the electron energy relaxation rate. Furthermore, we find the dependence of the time needed to evacuate the injected quasiparticles on the trap size. The evacuation time saturates at the lowest, diffusion-limited value upon extending the trap above a certain characteristic length $l_0$; the dependence of $l_0$ on the parameters of the trap and qubit is given in \eref{l0}.

The experimental findings reported in Sec.~\ref{sec:exp} validate the theoretical model. The relaxation rate $1/T_1$ of a transmon qubit is proportional to the quasiparticle density in the vicinity of the Josephson junction, making it possible to measure the dynamics of the quasiparticle population. We find that the population
decay rate increases with the length of the normal-metal traps, in agreement with the predicted cross-over from weak to strong trapping, see Fig.~\ref{fig:trapping_rate_fit}. For small traps we can estimate the effective trapping rate $\Gamma_\text{eff}$: both its order-of-magnitude and its increase with temperature indicate indeed a limitation due to the backflow of quasiparticles.

Utilizing traps is a viable strategy of mitigating the detrimental effect of quasiparticles on the qubits $T_1$ time. Further improvement of normal-metal traps may benefit from finding ways to shorten the electron energy relaxation time in them.

\acknowledgments

We gratefully acknowledge useful discussions with I. Khaymovich, J. Pekola, and C. Wang. This work was supported in part by the EU under REA Grant Agreement No. CIG-618258 (GC), DOE contract DEFG02-08ER46482 (LG), Max Planck award (RPR), and by ARO Grant W911NF-14-1-0011.

\appendix

\section{Tunneling rate equations.}\label{app:backflow_rate_eq}

In this Appendix, we derive the rate equations for quasiparticles and electrons accounting for tunneling between a superconductor and
a normal metal. Here we assume that both
the superconductor and the normal metal are sufficiently small volumes ($\Omega_S$ and $\Omega_N$, respectively),
such that the diffusion of excitations occurs on a fast time scale and the occupation probabilities are hence uniform in space.
Within these volumes we define the probabilities
\begin{eqnarray}
f\left(\xi_{m}\right) & = & \sum_{\sigma}\left\langle c_{m\sigma}^{\dagger}c_{m\sigma}\right\rangle \\
f_{\text{qp}}\left(\epsilon_{n}\right) & = & \sum_{\sigma}\left\langle \gamma_{n\sigma}^{\dagger}\gamma_{n\sigma}\right\rangle
\end{eqnarray}
of finding an electron excitation of energy $\xi_{m}$ in the
normal metal and a quasiparticle excitation of energy $\epsilon_{n}$
in the superconductor, respectively. The tunnel coupling between the
two, see Eq.~(\ref{eq:H_T}), gives rise to a change in both
occupation probabilities for energies above the gap, via processes whose rates can be computed using Fermi's Golden Rule:
\begin{eqnarray}
\dot{f}\left(\xi_{m}\right) & = & \sum_{n\sigma}\left[W_{n\sigma\rightarrow m\sigma}-W_{m\sigma\rightarrow n\sigma}\right .\\ \nonumber&& \left .+W_{0\rightarrow m\sigma,n-\sigma}-W_{m\sigma,n-\sigma\rightarrow0}\right],\\
\dot{f}_{\text{qp}}\left(\epsilon_{n}\right) & = & \sum_{m\sigma}\left[-W_{n\sigma\rightarrow m\sigma}+W_{m\sigma\rightarrow n\sigma}\right . \\ \nonumber&& \left .+W_{0\rightarrow m\sigma,n-\sigma}-W_{m\sigma,n-\sigma\rightarrow0}\right],
\end{eqnarray}
with
\begin{align}
W_{n\sigma\rightarrow m\sigma} = & \frac{2\pi}{\hbar}\frac{\left|\widetilde{t}\right|^{2}}{\Omega_S \Omega_N}u_{n}^{2}f_{\text{qp}}\left(\epsilon_{n}\right)\left[1-f\left(\xi_{m}\right)\right] \\ \nonumber & \times\delta\left(\epsilon_{n}-\xi_{m}\right)\, ,\\
W_{0\rightarrow m\sigma,n-\sigma} = & \frac{2\pi}{\hbar}\frac{\left|\widetilde{t}\right|^{2}}{\Omega_S \Omega_N}v_{n}^{2}\left[1-f_{\text{qp}}\left(\epsilon_{n}\right)\right] \\ \nonumber & \times\left[1-f\left(\xi_{m}\right)\right]\delta\left(\epsilon_{n}+\xi_{m}\right)\, .
\end{align}
The reverse processes are found by replacing $f_{(\qp)} \to 1 - f_{(\qp)}$.
Assuming particle-hole symmetry, $f\left(-\xi\right)=1-f\left(\xi\right)$,
we summarize the rate equations as
\begin{align}\label{eq:dot_f_N}
\dot{f}\left(\xi_{m}\right) = & \frac{2\pi}{\hbar}\frac{\left|\widetilde{t}\right|^{2}}{\Omega_S \Omega_N}\sum_{n}\left[f_{\text{qp}}\left(\epsilon_{n}\right)-f\left(\xi_{m}\right)\right]\delta\left(\epsilon_{n}-\xi_{m}\right)\\
\dot{f}_{\text{qp}}\left(\epsilon_{n}\right) = & \frac{2\pi}{\hbar}\frac{\left|\widetilde{t}\right|^{2}}{\Omega_S \Omega_N}\sum_{m}\left[f\left(\xi_{m}\right)-f_{\text{qp}}\left(\epsilon_{n}\right)\right]\delta\left(\epsilon_{n}-\xi_{m}\right). \label{eq:dot_f_S}
\end{align}

The tunneling processes considered above are elastic. In the
normal metal, for temperatures $T\ll\Delta$
there is a large interval of unoccupied states below the gap. Inelastic
processes, such as electron-phonon and electron-electron interactions,
can relax the excitations in the normal metal to energies below the gap, so that they cannot
return to the superconductor. We phenomenologically account for this relaxation by adding the term $-\Gamma_{\text{r}}f\left(\xi_{m}\right)$
to the right-hand side of Eq.~(\ref{eq:dot_f_N}). The relaxation rate $\Gamma_{\text{r}}$
is assumed energy-independent, which is justified if the interval
of non-zero excitations above the gap is within a narrow energy
strip of width $\ll\Delta$.

In the next step, we are interested in the probabilities to find excitations in the
states within a small energy interval $\delta\epsilon$. We define
the probability densities
\begin{align}
p_{N}\left(\epsilon\right) = & \frac{1}{N_S}\sum_{\epsilon<\xi_{m}<\epsilon+\delta\epsilon}f\left(\xi_{m}\right)\\
p_{S}\left(\epsilon\right) = &\frac{1}{N_S} \sum_{\epsilon<\epsilon_{n}<\epsilon+\delta\epsilon}f_{\text{qp}}\left(\epsilon_{n}\right)
\end{align}
which are normalized with respect to the normal-state number of states
in the superconductor $N_{S}= \nu_{S0} \Omega_S \delta \epsilon$. In the continuum limit $\delta\epsilon\rightarrow 0$ these definitions lead to \esref{pn_def} and \rref{ps_def}, respectively.
From \esref{eq:dot_f_N}-\rref{eq:dot_f_S} plus the phenomenological relaxation term discussed above, we obtain \esref{eq:dot_p_N}-\rref{eq:dot_p_S} with the rates
\begin{eqnarray}
\Gamma_{\text{esc}}\left(\epsilon\right) & = & \frac{2\pi}{\hbar}\left|\widetilde{t}\right|^{2}\frac{\nu_{S0}}{\Omega_{N}}\frac{\epsilon}{\sqrt{\epsilon^{2}-\Delta^{2}}}\, ,\\
\Gamma_{\text{tr}} & = & \frac{2\pi}{\hbar}\left|\widetilde{t}\right|^{2}\frac{\nu_{N0}}{\Omega_{S}} \, .
\end{eqnarray}

\section{Generalized diffusion equation and effective trapping rate}
\label{app:diffusion_eq}

In this Appendix we discuss the generalization of the rate equations \rref{eq:dot_p_N}-\rref{eq:dot_p_S} to include diffusion.
In disordered metals, the effect of elastic impurity scattering on the distribution function is accounted for by a diffusion term;
for quasiparticles in superconductor, the diffusion constant in the so-called ``hydrodynamical approach'' \cite{aggk} is energy-dependent:
\begin{align}
\nonumber & \dot{p}_{N}\left(\epsilon,\vec{r},t\right) =  D_{N}\vec{\nabla}^{2}p_{N}\left(\epsilon,\vec{r},t\right)+a\left(x,y\right)\delta\left(z-d_{S}\right)\\
& \times\left[\widetilde{\gamma}_{\text{tr}}p_{S}\left(\epsilon,\vec{r},t\right)
-\widetilde{\gamma}_{\text{esc}}\left(\epsilon\right)p_{N}\left(\epsilon,\vec{r},t\right)\right]\!-\Gamma_{\text{r}}p_{N}\left(\epsilon,\vec{r},t\right) , \label{diff_pn}\\ \nonumber
& \dot{p}_{S}\left(\epsilon,\vec{r},t\right) = D_{S}\left(\epsilon\right)\vec{\nabla}^{2}p_{S}\left(\epsilon,\vec{r},t\right)-a\left(x,y\right)\delta\left(z-d_{S}\right)\\
&\times\left[\widetilde{\gamma}_{\text{tr}}p_{S}\left(\epsilon,\vec{r},t\right)-\widetilde{\gamma}_{\text{esc}}\left(\epsilon\right)p_{N}\left(\epsilon,\vec{r},t\right)\right].
\label{diff_ps}
\end{align}
where $D_N$ is the diffusion constant in the normal-metal trap, and
\be
D_S (\epsilon) = D_S/\nu_S(\epsilon)
\ee
with $D_S$ being the normal-state diffusion constant in the superconductor; $\widetilde{\gamma}_\text{tr}$, $\widetilde{\gamma}_\text{esc}$, and $\Gamma_\text{r}$ are defined in Sec.~\ref{sec:diff}. The function $a(x,y)$ is
1 if coordinates $x$, $y$ belong to the normal-superconductor contact, 0 otherwise. For the $z$ coordinate, we assume that the superconductor occupies the interval $0< z < d_S$ and the interval $d_S < z < d_S+ d_N$ corresponds to the normal metal.

The above is a set of coupled linear differential equations that, for each energy $\epsilon$, can be in principle solved
in terms of eigenmodes, as done in Sec.~\ref{sec:dyn} for \eref{eq:diff_eq}.
Here, to justify that equation we consider the conditions under which \esref{diff_pn}-\rref{diff_ps} can be simplified, starting with the assumption that the normal metal is a thin layer.

\subsection{Thin normal metal}

In a sufficiently thin trap, the electron density
within the normal metal should not change significantly in $z$-direction.
This is the case if the length scale on which $p_{N}$ varies in the $z$ direction
is much larger than the thickness of the trap $d_N$.
Then we can expand $p_N$ as a function of the distance $z-d_N-d_S$ from the upper surface:
\[\label{pn_exp}\begin{split}
p_{N}\left(\epsilon,\vec{r},t\right)&=\frac{1}{d_N}\widetilde{p}_{N}\left(\epsilon,x,y,t\right)\\
&+\frac{\left(z-d_{N}-d_{S}\right)^{2}}{2d_{N}^{3}}p^{(2)}_{N}\left(\epsilon,x,y,t\right)+\ldots
\end{split}\]
The linear term is absent as to satisfy the boundary condition $\partial_{z}p_{N}=0$ at $z=d_{S}+d_{N}$. For the expansion to be applicable, we require $p^{(2)}_N \ll \widetilde{p}_N$; this condition will lead to a limit on the thickness $d_N$, as we show in what follows.

The diffusion equation \rref{diff_pn} for the normal metal may be alternatively expressed as
\[
\dot{p}_{N}\left(\epsilon,\vec{r},t\right)=D_{N}\vec{\nabla}^{2}p_{N}\left(\epsilon,\vec{r},t\right)-\Gamma_{\text{r}}p_{N}\left(\epsilon,\vec{r},t\right)
\]
for $z>d_{S}$, with the boundary condition at $z=d_{S}$
\[
\begin{split}
D_{N}\partial_{z}p_{N}\left(\epsilon,x,y,d_{S},t\right)+\widetilde{\gamma}_{\text{tr}}p_{S}\left(\epsilon,x,y,d_{S},t\right)\\
-\widetilde{\gamma}_{\text{esc}}\left(\epsilon\right)p_{N}\left(\epsilon,x,y,d_{S},t\right)=0\ .
\end{split}
\]
Using the Ansatz \rref{pn_exp}, from the two equations above we find in the leading order
\[\label{tp_n_d1}
\dot{\widetilde{p}}_{N}=D_{N}\vec{\nabla}^{2}\widetilde{p}_{N}+\frac{D_{N}}{d_{N}^{2}}p^{(2)}_{N}-\Gamma_{\text{r}}\widetilde{p}_{N}\ ,
\]
and
\[\label{pn2_lo}
-\frac{D_{N}}{d_{N}}p^{(2)}_{N}+d_N \widetilde{\gamma}_{\text{tr}}p_{S}-\widetilde{\gamma}_{\text{esc}}\left(\epsilon\right)\widetilde{p}_{N}=0\ .
\]
Solving \eref{pn2_lo} for $p_N^{(2)}$ and substituting it into \eref{tp_n_d1} we arrive at
\[\label{eq:diff_eq_thin_metal}
\dot{\widetilde{p}}_{N}=D_{N}\vec{\nabla}^{2}\widetilde{p}_{N}+
\widetilde{\gamma}_{\text{tr}}p_{S}-\Gamma_{\text{esc}}\left(\epsilon\right)\widetilde{p}_{N}
-\Gamma_{\text{r}}\widetilde{p}_{N}\ ;
\]
here the energy-dependent escape rate $\Gamma_\text{esc}(\epsilon)$ is defined by \esref{BCSdos} and \rref{eq_gamma_esc}.
We can formally solve this equation for $\widetilde{p}_{N}$ by introducing an appropriate set of eigenmodes, which are discrete due to the finite size of the normal metal.
Choosing for simplicity a rectangular trap with $0<x<d_x$ and $0<y < d_y$, we have eigenmodes of the form
\[
n(k_x,k_y)=\sqrt{N_x N_y}\cos(k_xx)\cos(k_y y)\ ,
\]
with $k_{x,y}=\pi n_{x,y}/d_{x,y}$, $n_{x,y}\in\mathbb{N}$, the normalisation constant $N_{x,y}=1+\delta_{0n_{x,y}}$, and the corresponding eigenvalues
\[\label{lambxy}
\lambda(k_x,k_y)=-D_Nk_x^2-D_Nk_y^2-\Gamma_\text{esc}(\epsilon)-\Gamma_\text{r}\ .
\]
We can now write the solution to \eref{eq:diff_eq_thin_metal} as
\[\label{eq:p_N_general}
\begin{split}
\widetilde{p}_N(x,y,t)= \Gamma_\text{tr}\int\frac{d\omega}{2\pi}\sum_{k_x,k_y}\frac{e^{-i\omega t}}{-i\omega-\lambda(k_x,k_y)}\\\times n(k_x,k_y)\tilde{p}_S(k_x,k_y,\omega)\ ,
\end{split}
\]
where we introduced the Fourier transform of $p_S$ at the interface,
\[\label{eq:p_S_Fourier}
\begin{split}
\tilde{p}_S(k_x,k_y,\omega)=\int_0^{d_x}\frac{dx}{d_x}\int_0^{d_y}\frac{dy}{d_y}n(k_x,k_y)\\ \times\int dt \, e^{i\omega t}p_S(x,y,d_S,t)\ .
\end{split}
\]
In the solution given in Eq.~\eqref{eq:p_N_general} we discarded any transient terms, which are exponentially suppressed for times $t\gg(\Gamma_\text{esc}(\epsilon)+\Gamma_\text{r})^{-1}$.

Let us assume that the length $L$ of the (largest) part of the superconductor not covered by the trap is sufficiently long (see end of Sec.~\ref{sec:etr}); then at the long times relevant to experiments, the time scale for the evolution of $p_S$ is determined by diffusion in the uncovered part. Similarly, diffusion in the region under the trap makes the long-time part of $p_S$ a smooth function of $x$ and $y$. This means that in \eref{eq:p_N_general} we can neglect all but the lowest mode and set $\omega=0$ and $k_x = k_y =0$ in the denominator. We thus arrive at
\[\label{pn_est}
\widetilde{p}_N \approx \frac{\widetilde{\gamma}_\text{tr} p_S}{\Gamma_\text{esc}(\epsilon)+\Gamma_\text{r}}\, .
\]
Using this estimate and the solution to \eref{pn2_lo}, the condition $p_N^{(2)} \ll \widetilde{p}_N$  can be written as a condition on the normal-metal thickness,
\[\label{dn_cond}
d_N\ll\sqrt{\frac{D_N}{\Gamma_\text{r}}}\ .
\]
Even for diffusion as slow as that of quasiparticles ($D_\qp > 10\,$cm$^2/$s \cite{wang}), using $\Gamma_\text{r} \sim 10^7\,{\rm s}^{-1}$ (see Sec.~\ref{sec:diff}), the right hand side is of order 10$\,\mu$m, much thicker than the film thickness in the experiments.

\subsection{Effective trapping rate}
\label{sec:etr}

We are interested in finding the equation governing the dynamics of $p_S$.
To this end, we substitute equations \rref{pn_exp} and \rref{pn_est} into \eref{diff_ps} to get at leading order the following equation:
\[\label{eq:app_diffusion_general}
\begin{split}
\dot{p}_{S}\left(\epsilon,\vec{r},t\right)&=D_{S}\left(\epsilon\right)\vec{\nabla}^{2}p_{S}\left(\epsilon,\vec{r},t\right)\\
&-a\left(x,y\right)\delta\left(z-d_{S}\right)\widetilde{\gamma}_{\text{eff}}\left(\epsilon\right)p_{S}\left(\epsilon,\vec{r},t\right)
\end{split}
\]
with
\[
\widetilde{\gamma}_{\text{eff}}\left(\epsilon\right)=\widetilde{\gamma}_{\text{tr}}\frac{\Gamma_{\text{r}}}{\Gamma_{\text{esc}}\left(\epsilon\right)+\Gamma_{\text{r}}}.
\]
Let us concentrate on the experimentally relevant case
of a thin superconductor in which $p_{S}$ varies slowly in the
$z$ direction with respect to the thickness $d_{S}$.
In analogy to the normal-metal case of the previous section, the $z$ dependence may
be neglected if $d_{S}\ll\sqrt{D_{S}\left(\epsilon\right)/\Gamma_{\text{eff}}\left(\epsilon\right)}$
with $\Gamma_{\text{eff}}=\widetilde{\gamma}_{\text{eff}}/d_{S}$.
Using the inequality
$D_{S}\left(\epsilon\right)/\Gamma_{\text{eff}}\left(\epsilon\right)>D_{S} \Gamma_{\text{esc}}/\Gamma_{\text{tr}}\Gamma_{\text{r}}$
(we remind that parameters $D_S$ and $\Gamma_\text{eff}$ without energy arguments correspond to the high-energy limiting values at $\epsilon \gg \Delta$),
we find the sufficient condition
\[
d_{S}\ll\sqrt{\frac{D_{S}\Gamma_{\text{esc}}}{\Gamma_{\text{tr}}\Gamma_{\text{r}}}}.
\]
For $d_{S}\approx d_{N}$ and $\nu_{S0}\approx\nu_{N0}$, this condition
reduces to $d_{S}\ll\lambda_S \equiv \sqrt{D_{S}/\Gamma_{\text{r}}}$ which, as discussed after \eref{dn_cond}, is generically satisfied in the experiments.
Then for a thin superconductor \eref{eq:app_diffusion_general} simplifies to
\[\label{eq:app_diffusion_2D}
\begin{split}
\dot{p}_{S}\left(\epsilon,x,y,t\right)& =D_{S}\left(\epsilon\right)\vec{\nabla}^{2}p_{S}\left(\epsilon,x,y,t\right)\\
&-a\left(x,y\right)\Gamma_{\text{eff}}\left(\epsilon\right)p_{S}\left(\epsilon,x,y,t\right).
\end{split}
\]
In closing this section, we note that the length scale $\lambda_S$ also determines the validity of the assumption made above that the uncovered part of the device is sufficiently long, which reads $L >\lambda_S$.

\subsection{Dynamics of the quasiparticle density}

The quantity that can be measured is the (normalized) quasiparticle density $x_\qp$, which is related to $p_S$ by integration over energy,
$x_{\text{qp}}=\frac{2}{\Delta}\int_{\Delta}^{\infty}d\epsilon\,  p_{S}\left(\epsilon\right)$.
Focusing on the case of thin films,
using \eref{eq:app_diffusion_2D} as reference we write
\[\label{xpeq}
\dot{x}_\text{qp}=D_\text{qp}\vec{\nabla}^2 x_\text{qp}-a(x,y){\Gamma}_\text{eff}x_\text{qp}
\]
where $x_\text{qp}$ depends only on $x$ and $y$. Comparing this equation to the integral over energy of \eref{eq:app_diffusion_2D} we identify the coefficient $\Gamma_\text{eff}$ with
\[\label{geffph}
\Gamma_{\text{eff}}=\frac{\int_{\Delta}^{\infty}d\epsilon \, \Gamma_{\text{eff}}\left(\epsilon\right)p_{S}\left(\epsilon\right)}{\int_{\Delta}^{\infty}d\epsilon \, p_{S}\left(\epsilon\right)}
\]
[cf. \eref{eq:Geffdef}], and an analogous relation holds  between the quasiparticle diffusion coefficients $D_\qp$ and $D_S(\epsilon)$. These relations are exact if the energy dependence in $p_S$ can be factorized from its temporal and spatial ones.
Although the structure of \esref{eq:app_diffusion_general} and \rref{eq:app_diffusion_2D} does not favor such factorization, the latter may be enforced by the process of quasiparticle thermalization due to their interaction with phonons. The temperature dependence of $\Gamma_{\rm eff}$ quoted in the text after \eref{eq:Geffdef} assumes the quasiparticles are thermalized at an effective temperature $T$.
Phenomenological equations such as \esref{xpeq}-\rref{geffph} are widely used in the literature \cite{ullom,plourde,wang,raja1,raja2,ullom2} as they successfully describe experiments as in Sec.~\ref{sec:exp}.


\begin{thebibliography}{30}
\bibitem{note_tL}We note that while this result for $\tau_w$ was derived assuming a short trap, $d \ll \lambda_\text{tr}$, it is more generally valid even for longer trap with $d>\lambda_\text{tr}$ as long as we identify $L$ with the longest section of the wire not covered by the normal metal, and the condition $L \gg \lambda_\text{tr}$ is satisfied. The effects of a not-small $d/\lambda_\text{tr}$ ratio and of the device geometry will be considered in more detail elsewhere.
\bibitem{notefit}While these fitted values for $t_L$ are in reasonable agreement with the values estimated using \eref{tL}, we note that in general they can be affected by details of the device geometry and decay measurement process. We will return on these aspects elsewhere.
\end{thebibliography}
\end{document}